# Numerical simulation and analytical modelling of self-heating in FDSOI MOSFETs down to very deep cryogenic temperatures

G. Ghibaudo[1], M. Cassé[2] and F. Balestra[1]

1) IMEP-LAHC, Univ. Grenoble Alpes, Minatec, 38016 Grenoble, France.
2) LETI-CEA, Univ. Grenoble Alpes, Minatec, 38054 Grenoble, France.
Email: gerard.ghibaudo@gmail.com.

**Abstract**

Self-heating (SHE) TCAD numerical simulations have been performed, for the first time, on 30nm FDSOI MOS transistors at extremely low temperatures. The self-heating temperature rise $dT_{max}$ and the thermal resistance $R_{th}$ are computed as functions of the ambient temperature $T_a$ and the dissipated electrical power ($P_d$), considering calibrated silicon and oxide thermal conductivities. The characteristics of the SHE temperature rise $dT_{max}(P_d)$ display sub-linear behavior at sufficiently high levels of dissipated power, in line with standard FDSOI SHE experimental data. It has been observed that the SHE temperature rise $dT_{max}$ can significantly exceed the ambient temperature more easily at very low temperatures. Furthermore, a detailed thermal analysis of the primary heat flows in the FDSOI device has been conducted, leading to the development of an analytical SHE model calibrated against TCAD simulation data. This SHE analytical model accurately describes the $dT_{max}(P_d)$ and $R_{th}(T_a)$ characteristics of an FDSOI MOS device operating at extreme low ambient temperatures. These TCAD simulations and analytical models hold great promise for predicting the SHE and electro-thermal performance of FDSOI MOS transistors against ambient temperature and dissipated power.

# 1. Introduction

Cryogenic electronics remains an important area of research due to its ability to enhance circuit performance in terms of operation speed, turn-on behavior, thermal noise reduction, and punch-through decrease [1-5]. It finds applications in various fields such as high-speed computing, detection and sensing, spatial electronics, readout CMOS electronics, and q-bit MOS devices for quantum computing [6-8]. However, there are still several challenges in characterizing and modeling CMOS devices at very low temperatures. Recent efforts have focused on achieving TCAD device simulations at deep cryogenic temperatures [9-11], with a particular emphasis on conducting electro-thermal simulations to assess self-heating (SHE) in FDSOI MOS devices at low temperatures [12-16]. However, self-heating electro-thermal simulations in CMOS devices at deep cryogenic temperatures are still lacking due to the significant challenges involved. Such simulations could provide crucial insights into the electro-thermal properties and self-heating phenomena in FDSOI MOS devices operating within the sub-Kelvin temperature range, especially in the context of quantum computing.

Therefore, in this study, building upon our previous work dedicated to electro-thermal transport coefficients in FDSOI MOS devices operating under deep cryogenic conditions [17], we present, for the first time, self-heating 2D TCAD semi-classical simulations performed on 30nm FDSOI MOS transistors at very low temperatures (around 10mK). We investigate the temperature rise caused by self-heating and evaluate the thermal resistance as functions of ambient temperature and dissipated power, taking into account calibrated material thermal conductivity. Additionally, by conducting a detailed thermal analysis of the primary heat flows within the FDSOI structure, we develop an analytical model for self-heating that accurately describes the temperature rise and thermal resistance in FDSOI MOS devices across a range of dissipated powers, even when operating at ambient temperatures as low as 10mK. This comprehensive approach allows us to predict the electro-thermal performance of FDSOI MOS transistors in extremely deep cryogenic conditions.

# 2. Theoretical background and simulation methodology

In this section, we will discuss the equations governing electro-thermal transport in the FDSOI structure illustrated in Fig. 1. The simulated FDSOI structure comprises a 4nm top oxide thickness, a 6nm undoped silicon film, and a 20nm bottom oxide (BOX), with a channel length of 30nm. To simplify the computations, we have omitted the silicon substrate beneath the BOX and the physical source and drain regions.

The transport parameters, such as carrier density ($n$), electron mobility ($\mu$), electronic conductivity ($\sigma$), thermopower ($S$), and electronic thermal conductivity ($K_e$), are formulated using analytical approximations suitable for 2D TCAD semi-classical simulations. Furthermore, empirical analytical functions, calibrated using literature experimental data, are employed to express the lattice thermal conductivity in the silicon channel, as well as in the top and bottom oxides.

Finally, we present the methodology for solving the Poisson and electro-thermal transport equations, along with the associated boundary conditions. It should be mentioned that herein specific boundary conditions are applied on a single 30nm channel length device in order to properly differentiate short and long channel situations.

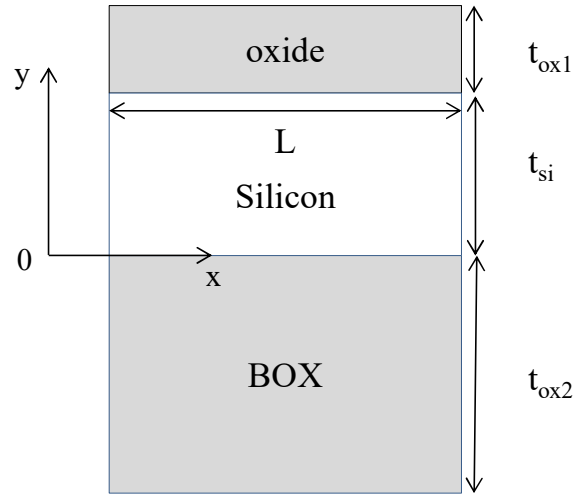

**Fig. 1.** Schematic of FDSOI structure used for simulation ($t_{ox1}$=4nm, $t_{ox2}$=20nm, $t_{si}$=6nm, L=30nm).

2.1. Electro-thermal transport equations

In silicon, the equations governing electron charge and heat transport can be expressed using linear response theory as follows [18]:

$$J_e = \sigma . \nabla U_c - \sigma . S . \nabla T \tag{1a}$$

$$J_q = \sigma . T . S . \nabla U_c - K_{th} . \nabla T \tag{1b}$$

where $J_e$ is electronic current density (A/cm$^2$), $J_q$ is the heat thermal flux density (W/cm$^2$), $U_c$ is the electron quasi-Fermi level, $T$ is the local lattice temperature, $\sigma$ is the electronic conductivity, $S$ is the thermopower (or Seebeck coefficient) and $K_{th}$ is the heat thermal conductivity.

The local electrical conductivity $\sigma$ in the silicon channel is evaluated by considering a standard mobility law both depending on temperature, local vertical electric field $F_y$ and lateral electric field $F_x$ as [11, 17, 19],

$$\sigma(T) = q . \mu . n \tag{2a}$$

with
$$\mu_0(T) = \left[\frac{1}{5000} + \frac{1}{600}\left(\frac{T}{300}\right)\right]^{-1} (cm^2/Vs) \tag{2b}$$

and
$$\mu = \frac{\mu_0(T)}{\left[1+\left(\frac{F_y}{F_{cy}}\right)^2\right]\left(1+\frac{F_x}{F_{cx}}\right)} \tag{2c}$$

where $F_{cy}$ is a critical field ($\approx$1MV/cm) allowing to emulate the mobility degradation at high gate voltage caused by surface roughness scattering, whereas $F_{cx}=v_{sat}/\mu_0$, with $v_{sat}$ [$=1.3\times10^7/(1+T/1000)$] being the electron saturation velocity (cm/s), accounting for the mobility decrease due to saturation velocity effect in short channel devices. Note that the latter effect is disabled in long channel case. The low field empirical mobility law $\mu_0(T)$ describes the mobility increase with temperature reduction governed by phonon scattering and for mobility saturation at low temperature due to prevailing defective scattering [2, 20]. It should be mentioned that the effective mobility $\mu_{eff}$ deduced from this local mobility law $\mu$ of Eq. (2c) exhibits variation with temperature in agreement with experimental data obtained on FDSOI MOSFETs [20, 21].

The electron density $n$ is evaluated through an analytical approximation of the Fermi-Dirac function valid from Boltzmann to metallic statistics and given by [11, 17],

$$n(u_f) = N_C(T) \cdot \left[\ln\left(1 + e^{\frac{2u_f}{3}}\right)\right]^{3/2} \cdot \left[1 + p + p \cdot \tanh\left(\frac{u_f+2}{2}\right)\right] \quad \text{with } p\approx 0.195 \quad (3)$$

where $u_f = q(V - V_0 - U_c)/kT$ is the reduced Fermi energy, $V$ being the local electric potential, $V_0$ ($\approx$0.55V) the mid-gap potential and $N_C(T)$ the effective density of states ($= 2 \times 10^{19} \cdot \left(\frac{T}{300}\right)^{\frac{3}{2}}$ for silicon). It should be noted that Eq. (3) provides a carrier density value with an error inferior to $\approx$5% over the whole reduced Fermi energy range as compared to the exact Fermi-Dirac function [17].

Similarly, the thermopower $S$ is computed as function of the reduced Fermi level with the analytical approximation accounting for the Fermi-Dirac statistics and given by [17],

$$S(u_f) = \frac{k}{q} \cdot \left\{\left[\frac{1}{2+\ln\left(1+\frac{N_C(T)+4 \cdot n(u_f)}{n(u_f)}\right)}\right]^a + \left[\frac{1}{\frac{\pi^2}{2} \cdot \left[\frac{4}{3\sqrt{\pi}} \cdot \frac{N_C(T)}{n(u_f)}\right]^{2/3}}\right]^a\right\}^{-1/a} \quad \text{with } a\approx 2. \quad (4)$$

Equation (4) offers a reliable description of the thermopower, exhibiting an error of approximately 7% or less across the entire reduced Fermi energy range when compared to the exact solution obtained using Fermi-Dirac statistics [17]. However, it is important to note that in this self-heating simulation study, we have neglected the contribution of phonon drag to the thermopower, as discussed in [17]. It is acknowledged that further research is required to incorporate this phenomenon into the electro-thermal transport equations, but such an endeavor extends beyond the scope of this paper.

The heat thermal conductivity $K_{th}$ in silicon consists of two components: one due to heat conduction by phonon diffusion in bulk silicon ($K_{si}$), and the other from the electronic thermal conductivity of the electron gas in the channel ($K_e$). Therefore, we have $K_{th}=K_{si}+K_e$. The electronic thermal conductivity ($K_e$) can be accurately approximated by an analytical expression that is valid across the semiconductor to metallic regime and which is given by [17],

$$\frac{K_e}{\sigma.T}(u_f) = \left[\left(S(u_f)^2\right)^b + \left(\frac{\pi^2}{2}.\left(\frac{k}{q}\right)^2\right)^b\right]^{1/b} \qquad \text{with } b \approx 1.1. \qquad (5)$$

Eq. (5) also offers a good approximation to the exact results with an overall error below ≈10% over the whole Fermi energy range [17].

It is worth noting that the electrical conductivity in the top and bottom oxides is considered to be zero, as they are treated as perfect insulating dielectrics.

The bulk silicon heat conductivity data from Callaway [22] were adjusted to consider the reduction of $K_{si}$ in thin silicon films (in this case, 6nm) caused by increased phonon boundary scattering. This adjustment was performed using the approach proposed by Ashegi et al. [23]. The data were then interpolated and extrapolated with respect to temperature using a combined power law analytical function of the form [17],

$$K_{si}(T) = (1.06 \times 10^4.T^{-1.7} + 2.6 \times 10^{-2}.T)^{-1} \quad [\text{W/(cm.K)}]. \qquad (6)$$

Since the silicon channel is surrounded by a dielectric material (silicon dioxide), it is important to consider the heat conduction in the silicon dioxide as well. For this purpose, the heat conductivity data for silicon dioxide ($K_{ox}$) adapted from Lee and Cahill [24] were fitted using a combined power law analytical function of the form,

$$K_{ox}(T) = (1.5 \times 10^4.T^{-1} + 6.3 \times 10^{-2}.T)^{-1} \quad [\text{W/(cm.K)}]. \qquad (7)$$

As can be seen from Fig. 2, these fitting functions provide good approximations for both $K_{si}$ and $K_{ox}$ as a function of temperature, especially in the cryogenic range where they will be extrapolated.

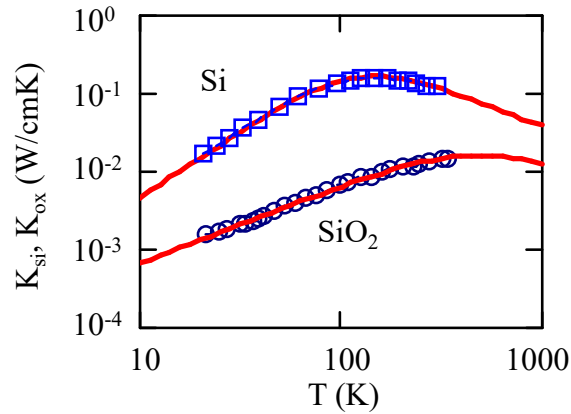

**Fig. 2.** Experimental (symbols) and fitted (solid lines) variations of heat thermal conductivity for silicon $K_{si}$ and silicon dioxide $K_{ox}$ as a function of temperature T. Experimental data adapted from Refs [22, 24].

2.2. Numerical simulation methodology

To simulate the self-heating in the FDSOI n MOSFET shown in Fig. 1, it is necessary to solve the Poisson equation, the electronic current continuity equation, and the heat flow continuity equation self-consistently. By solving these equations simultaneously, the electro-thermal behavior of the device can be accurately modeled, taking into account the interaction between electrical and thermal effects.

The Poisson equation governing the electrical potential $V$ in the FDSOI structure with channel residual doping level $N_a$ reads,

$$\nabla(\varepsilon_{si} \nabla V) = q \cdot \left[ n\left(\frac{q(V-V_0-U_c)}{kT}\right) - p\left(\frac{q(-V-V_0)}{kT}\right) - \frac{N_a}{1+4.exp\left(\frac{q(U_c-V+V_a)}{kT}\right)} \right] \quad \text{(Silicon)} \quad (8a)$$

$$\nabla(\varepsilon_{ox} \nabla V) = 0 \quad \text{(Oxide)} \quad (8b)$$

where $\varepsilon_{ox}$ and $\varepsilon_{si}$ being the oxide and silicon permittivities, respectively, $V_0$ ($\approx$0.55V) the mid-gap potential and $V_a$ the impurity ionization potential ($V_a=V_0-50$mV). The carrier concentrations in Eq. (8a) were computed using the analytical approximation of Eq. (3). In practice, the hole concentration was overlooked as being negligible in an n MOSFET structure biased from weak to strong inversion.

The electronic current continuity equation in the silicon channel is given by,

$$\nabla(J_e) = \nabla(\sigma \cdot \nabla U_c - \sigma \cdot S \cdot \nabla T) = 0 \quad \text{(Silicon)} \quad (9)$$

where $\sigma$ and $S$ are evaluated with Eqs (2) and (4).

Similarly, the heat thermal flow continuity equations in the FDSOI structure accounting for the Joule dissipation in the silicon channel are expressed as,

$$\nabla(J_q) = \nabla(\sigma \cdot T \cdot S \cdot \nabla U_c - K_{th} \cdot \nabla T) = |J_e|^2/\sigma \quad \text{(Silicon)} \quad (10a)$$

$$\nabla(J_q) = \nabla(-K_{ox} \cdot \nabla T) = 0 \quad \text{(Oxide)} \quad (10b)$$

where $K_{th}=K_{si}+K_e$ is given by Eqs (5) and (6) and $K_{ox}$ is given by Eq. (7). Note the presence of the source term $|J_e|^2/\sigma$ in Eq. (10a) responsible for the self-heating in the silicon channel.

The equations (8) to (10) mentioned above were solved self-consistently in two dimensions using the finite element software FlexPDE [25]. Proper boundary conditions (BC) were defined within the FDSOI structure for the numerical simulations. The boundary conditions utilized in the simulations are summarized in Fig. 3 for the two thermal BC configurations considered in this study. In both cases, the electrical BC's are identical. Note that here specific boundary conditions are applied on a 30nm channel length device in order to differentiate short and long channel situations. The electrical potential V is fixed to the front gate voltage $V_{g1}$ at the top gate (G) and to zero at the bottom gate (B). The quasi-Fermi potential $U_c$ is set to zero at the source (S) and to the drain voltage $V_d$ at the drain (D). At the source and drain boundaries, $dV/dx=0$ is imposed to emulate a long channel condition. In the case of a short channel, V is fixed to $V_{bi}$ at the source and to ($V_{bi}+V_d$) at the drain, where $V_{bi} \approx$ 0.6V for a degenerate source-drain region with an n-type doping level $N_d=5\times10^{19}$/cm$^3$. In the "BSDG" BC configuration, the temperature $T$ is fixed to the ambient temperature $T_a$ at all boundaries, including the front gate (G), source (S), drain (D), and bottom gate (B). This configuration represents the case where the FDSOI device is thermally grounded at all electrodes, providing the best thermalization scenario. In the "B" BC configuration, the temperature $T$ is only fixed to $T_a$ at the bottom gate (B), assuming no or negligible heat flow at the source, drain, and top gate electrodes. This configuration represents the worst thermalization case since the bottom oxide (BOX) has the highest thermal resistance. It should be noted that this "B" BC configuration does not account for the thermal resistance of the silicon substrate, which is omitted in the simulation. However, based on separate heat transport simulations, it has been found that the thermal spreading resistance of a thick silicon layer (300 μm) underneath a narrow BOX (30nm) is negligible compared to the BOX thermal resistance due to the lower thermal conductivity of the BOX material (see Fig. 2).

Figure 4 illustrates the typical results obtained from numerical simulations for the temperature rise ($dT=T-T_a$) with the "BSDG" BC configuration, where the ambient temperature ($T_a$) is set to 10K. It can be observed that $dT$ exhibits a bell-shaped behavior with a peak located in the middle of the silicon film near the front oxide interface, where the channel is formed. This behavior is obtained by applying a front gate voltage $V_{g1}$=1V and a drain voltage $V_d$=43mV. In the presented example, the maximum temperature rise ($dT_{max}$) is approximately 12K, which exceeds the imposed ambient temperature $T_a$ of 10K at all boundaries in the "BSDG" configuration. It should be mentioned that as $V_d$ is increased, the device goes into triode regime and the maximum temperature rise $T_{max}$ is slightly displaced towards the drain side. In any case, once the maximum temperature rise is determined, the effective thermal resistance ($R_{th}$) can be calculated using the following formula:

$$R_{th} = \frac{dT_{max}}{P_d} \qquad (11)$$

where $P_d=I_d.V_d$ is the electrical power dissipated in the FDSOI device, $I_d$ being the drain current. In the example of Fig. 4, $R_{th}\approx$38K/W. Therefore, Eq. (11) will be used in the next section to calculate $R_{th}$ as functions of power and ambient temperature for both "BSDG" and "B" BC configurations for the FDSOI MOSFET of Fig. 1.

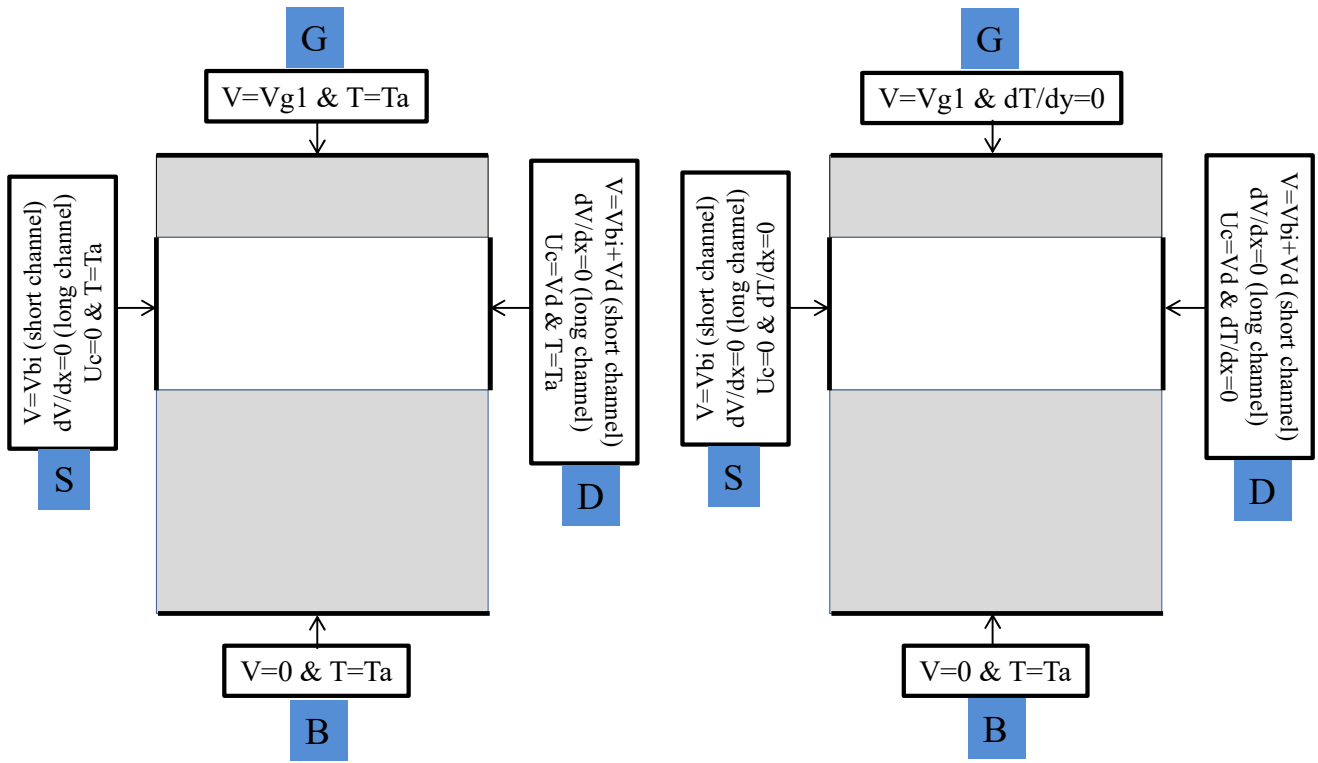

**Fig. 3.** Schematic showing the two typical boundary conditions used for the numerical simulations carried out this work.

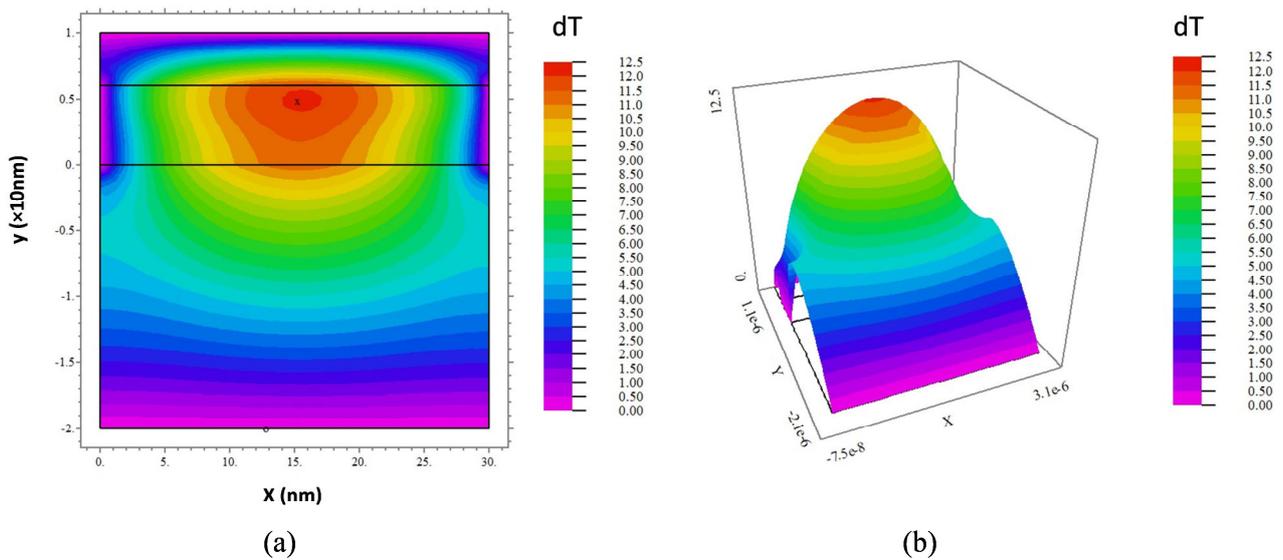

(a)                                                (b)

**Fig. 4.** Example of contour plot (a) and surface plot (b) of temperature rise $dT$ obtained by numerical simulation in a FDSOI MOSFET (parameters: $V_{g1}$=1V, $P_d$=0.31W, $T_a$=10K).

## 3. Results and discussion

Prior to simulations including self-heating, drain current transfer characteristics were simulated in linear operation ($V_d$=1mV) with only Eq. (9) i.e. without self-heating consideration in order to illustrate their general temperature dependence and adequacy with experimental behavior.

Figure 5 shows typical $I_d(V_{g1})$ transfer characteristics in linear and log scale obtained for various temperatures $T=T_a$ by numerical simulation without self-heating in a FDSOI MOSFET with long channel BC (see Fig. 3). One should notice first the significant increase of the drain current at strong inversion as the ambient temperature is lowered due to the huge improvement of the low field mobility (see Eq. (2b)). The presence of a zero-temperature-coefficient point is also noticeable and well in line with experimental results [2, 20, 26]. One should also note the strong decrease with ambient temperature of the subthreshold swing SS ($=dV_{g1}/dlog(I_d)$) associated to the thermal energy reduction. This can be better seen in Fig. 6 where the subthreshold swing SS is plotted versus ambient temperature $T_a$ in linear (left) and log scale (right). It is worth noting that here SS is saturating at low temperature below ≈30K, well mimicking typical experimental data [20, 27]. This behavior has been mainly interpreted by the presence of localized-states exponential band tails due to potential-fluctuations-induced disorder [27-29]. In our numerical simulations, this phenomenon has been taken into account by considering an effective statistical temperature $T_{eff}$, introduced in Eq. (3) defining the carrier concentration, and given by [29],

$$T_{eff} = T_S \cdot \left[ 1 + \alpha \cdot \ln\left( 1 + e^{\frac{T-T_S}{\alpha \cdot T_S}} \right) \right] \qquad (12)$$

where $kT_S$ is the energy width of the band tails and $\alpha$=0.2 is an empirical parameter. By considering the effective statistical temperature $T_{eff}$ in the simulations, we can effectively capture the influence of exponential band tails in semi-classical TCAD simulations.

Additionally, it is important to note that the simulated $I_d(V_{g1})$ transfer characteristics shown in Fig. 5 and the corresponding subthreshold swing SS depicted in Fig. 6 are in good agreement with experimental data obtained from FDSOI MOSFETs at cryogenic temperatures [2, 5, 20, 30]. This agreement further confirms the validity of our simulation assumptions and supports the reliability of our modeling approach in capturing the essential device behavior under cryogenic conditions.

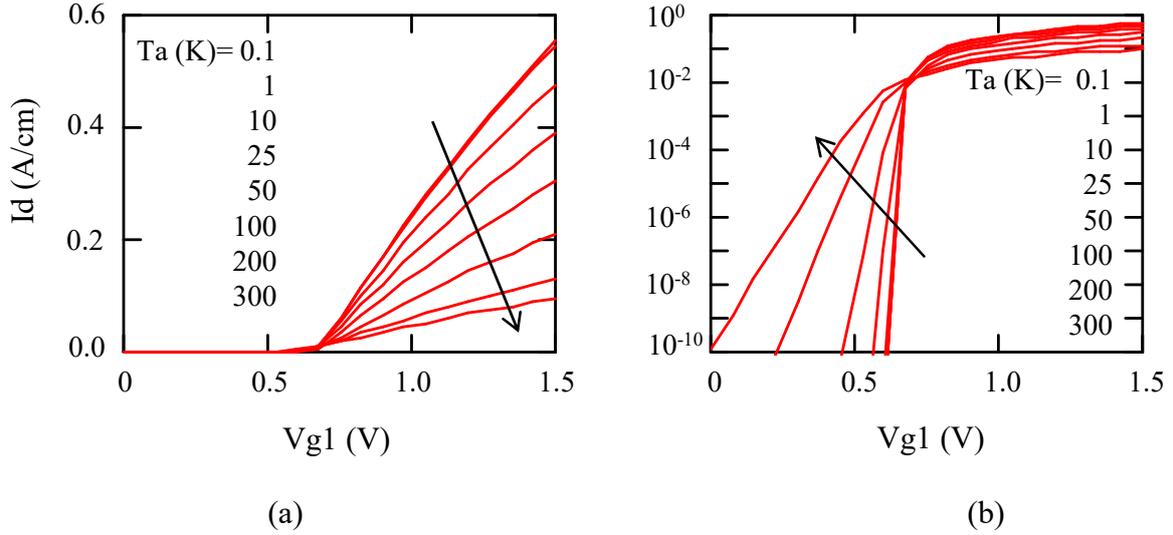

**Fig. 5.** Temperature dependence of the transfer characteristics $I_d(V_{g1})$ in linear (a) and log (b) scales as obtained by numerical simulation without considering the self-heating in a FDSOI MOSFET for long channel case ($t_{ox1}$=4nm, $t_{ox2}$=10nm, $t_{si}$=6nm, L=30nm, $V_d$=1mV, $T_S$=25K).

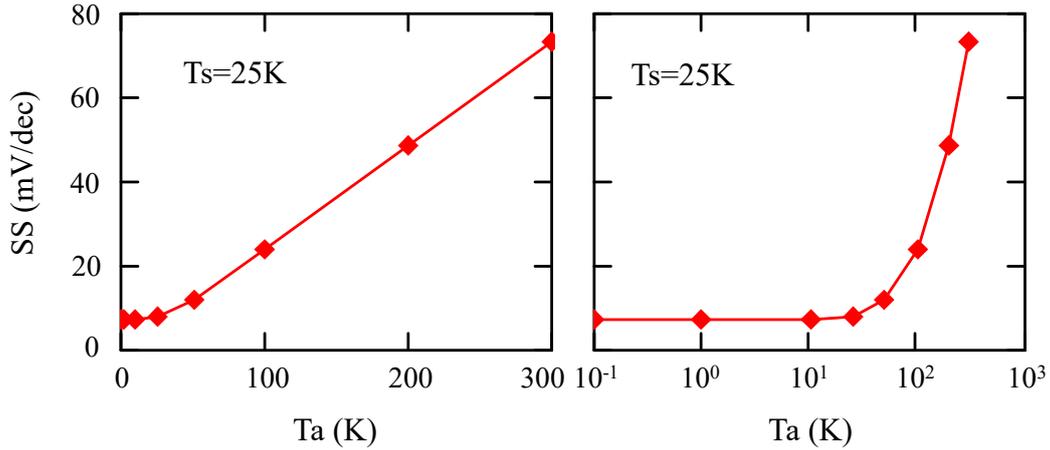

**Fig. 6.** Temperature dependence of the subthreshold swing SS in linear scale (left) and log scale (right) as obtained from the transfer curves of Fig. 5 (long channel case FDSOI MOSFET, $t_{ox1}$=4nm, $t_{ox2}$=10nm, $t_{si}$=6nm, L=30nm, $V_d$=1mV, $T_S$=25K).

### 3.1. Self-heating results for "BSDG" BC configuration

We conducted simulations of typical output drain current $I_d(V_d)$ characteristics for a long channel BC FDSOI MOSFET under various ambient temperatures, both with and without self-heating effects (see Fig. 7a). The corresponding relative changes in drain current due to self-heating, $dI_d/I_d$, are presented in Fig. 7b. It is important to highlight the significant influence of self-heating on the amplitude of the drain current, particularly as the ambient temperature is decreased towards deep cryogenic conditions. These results underscore the importance of considering self-heating effects when analyzing the performance of FDSOI MOSFETs operating in extreme temperature environments.

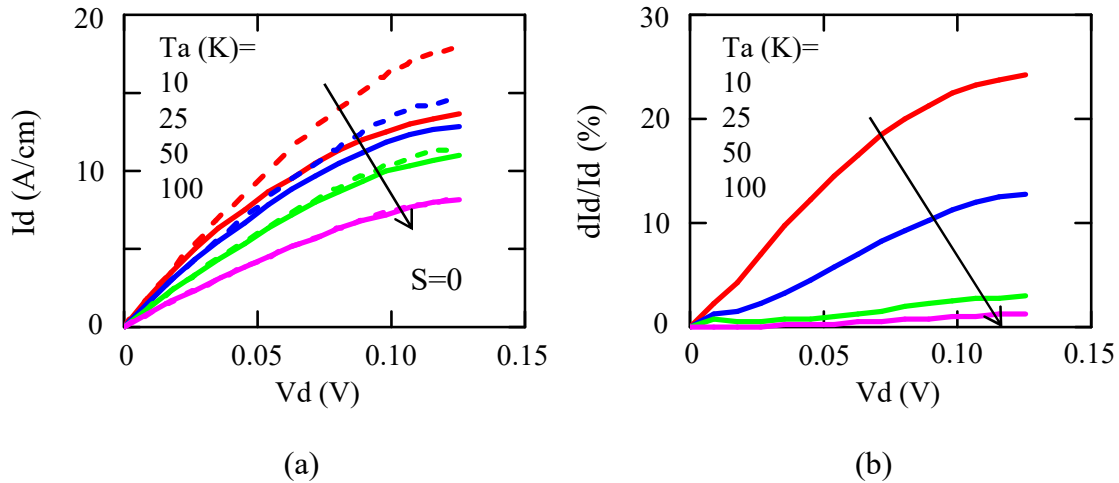

(a)                 (b)

**Fig. 7.** a) Temperature dependence of (a) $I_d(V_d)$ output characteristics obtained by numerical simulations with (solid lines) and without (dashed lines) self-heating.

(b) Corresponding relative drain current change $dI_d/I_d(V_d)$ due to self-heating (long channel case FDSOI MOSFET, $t_{ox1}$=4nm, $t_{ox2}$=10nm, $t_{si}$=6nm, L=30nm, $V_{g1}$=1V, $T_S$=25K).

Figure 8 presents a comparison of the drain current output characteristics, $I_d(V_d)$, obtained from numerical simulations without (a) and with (b) consideration of the thermopower ($S$) term in the electro-thermal transport equations (1). Upon initial observation, it appears that the inclusion of the thermopower term has a negligible impact on the drain current characteristics.

Similarly, in Fig. 9, the maximum temperature rise characteristics, $dT_{max}(P_d)$, are presented as function of power $P_d$, as obtained from numerical simulations without (a) and with (b) the inclusion of the thermopower in the electro-thermal transport equations. It can be observed that the influence of the thermopower on the maximum temperature rise appears to be minor, particularly at deep cryogenic temperatures. This observation is further supported by the plot of the corresponding thermal resistance $R_{th}$ as function of power $P_d$, as shown in Fig. 10, where only slight differences are observed, mainly at 100K. This result can be attributed to the fact that the thermopower in the metallic regime exhibits a linear variation with temperature and, as a consequence, diminishes at cryogenic temperatures [17].

Based on the observed characteristics and results, it can be concluded that, as a first approximation, the thermopower term can be neglected in the electro-thermal transport equations for self-heating numerical simulations. This simplification is valuable as it significantly reduces the computation time, typically by a factor of 4-5, by more decoupling the electro-thermal transport equations (1a) and (1b). Therefore, in most of our self-heating studies, we will primarily focus on numerical results obtained with $S=0$, assuming no thermopower contribution.

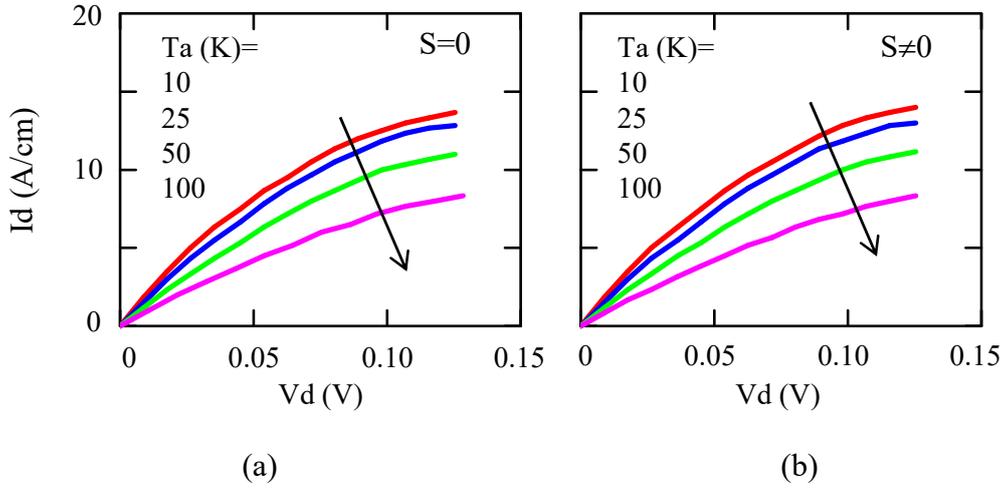

(a)  (b)

**Fig. 8.** Temperature dependence of $I_d(V_d)$ output characteristics obtained by numerical simulations without (a) and with (b) consideration of thermopower in the electro-thermal transport equations (long channel case FDSOI MOSFET, $t_{ox1}$=4nm, $t_{ox2}$=10nm, $t_{si}$=6nm, L=30nm, $V_{g1}$=1V, $T_S$=25K).

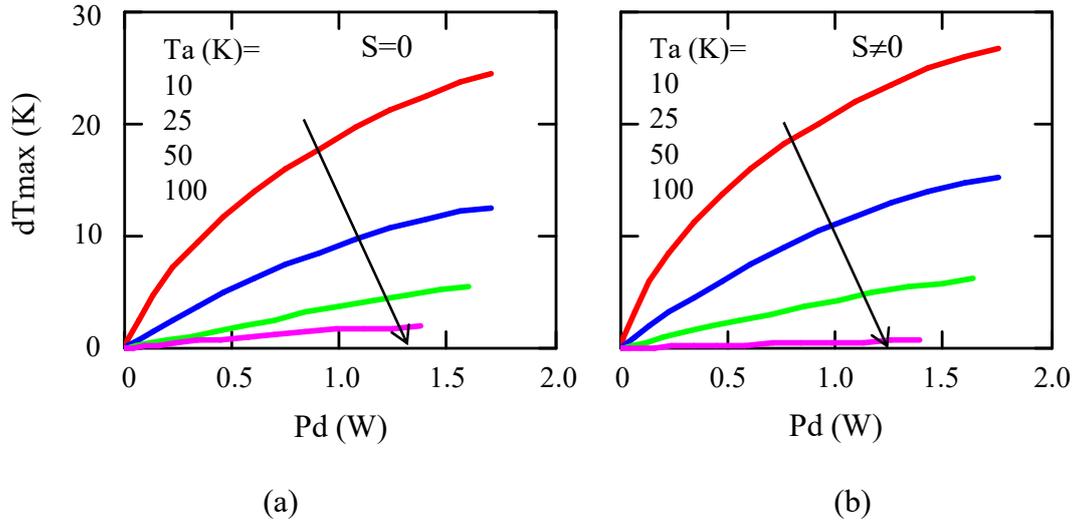

(a)  (b)

**Fig. 9.** Temperature dependence of $dT_{max}(P_d)$ characteristics obtained by numerical simulations without (a) and with (b) consideration of thermopower in the electro-thermal transport equations (long channel case FDSOI MOSFET, $t_{ox1}$=4nm, $t_{ox2}$=10nm, $t_{si}$=6nm, L=30nm, $V_{g1}$=1V, $T_S$=25K).

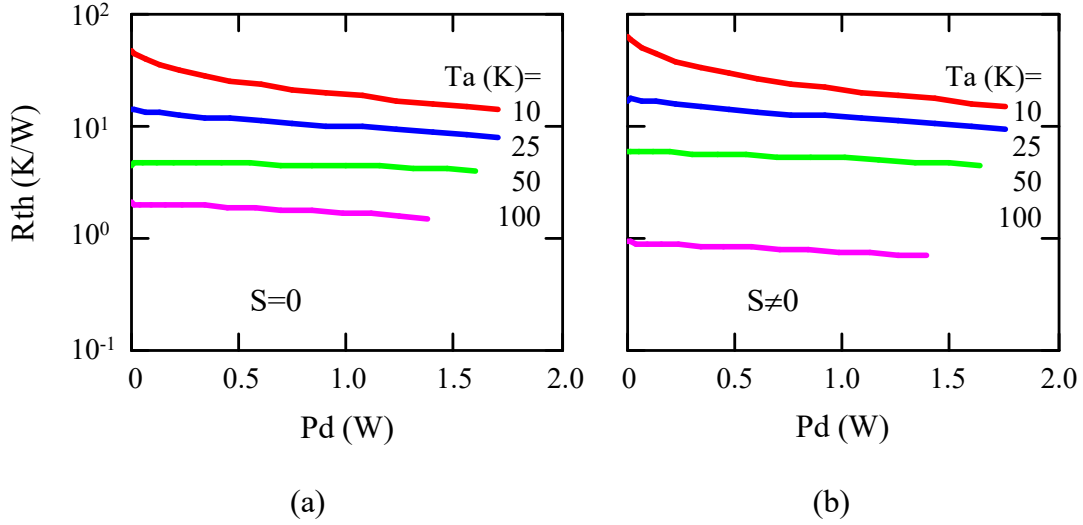

**Fig. 10.** Temperature dependence of $R_{th}(P_d)$ characteristics obtained by numerical simulations without (a) and with (b) consideration of thermopower in the electro-thermal transport equations (long channel case FDSOI MOSFET, $t_{ox1}$=4nm, $t_{ox2}$=10nm, $t_{si}$=6nm, L=30nm, $V_{g1}$=1V, $T_S$=25K).

In Figure 11a, the variations of the thermal resistance $R_{th}$ with the ambient temperature $T_a$ are shown for various drain voltage values. The solid lines represent the results obtained by numerical simulations without considering the thermopower term in the electro-thermal transport equations, while the symbols represent the results with the inclusion of the thermopower term. These results further confirm that the thermopower term can be mostly neglected in the self-heating electro-thermal simulations. It is noteworthy that as the drain voltage increases, the thermal resistance $R_{th}$ saturates at an earlier ambient temperature, while still following the same decreasing asymptotic trend with increasing ambient temperature.

In Figure 11b, a comparison is shown between the variations of thermal resistance $R_{th}(T_a)$ obtained from self-heating numerical simulations and semi-analytical modeling. The analytical model is based on the equivalent circuit of thermal resistances depicted in Fig. 12. To construct the analytical model, it is first noted that the one-dimensional thermal resistance $R_{1D}$ of a material with conductivity $K_1(T)$, length $L_1$, and width $W_1$, subjected to a temperature difference $dT$, can be expressed as Equation (A1), indicating that $R_{1D}$ is a function not only of temperature $T$ but also of $dT$. Considering the "BSDG BC" condition, where the thermal heat primarily flows from the hot point in the channel to the gate through the front oxide, to the body through the BOX, and to the source and drain along the silicon channel, the global equivalent resistance $R_{thBSDG}(T,dT)$ can be derived from the heat flow partition as expressed in Eq. (A6). The vertical resistance $R_{sit}$ is neglected due to the thinness of the silicon layer. The modeling results in Fig. 12b are obtained from $R_{thBSDG}(T_a,dT_{max})$, where $dT_{max}$ corresponds to the values obtained from the numerical simulations for $R_{th}(T_a)$. With Equation (A6), the saturation of $R_{th}$ at low $T_a$ for higher $V_d$ values can now be interpreted as the result of the strong nonlinearity of $R_{thBSDG}(T_a,dT_{max})$ due to larger $dT_{max}$ at higher power i.e. higher $V_d$ values.

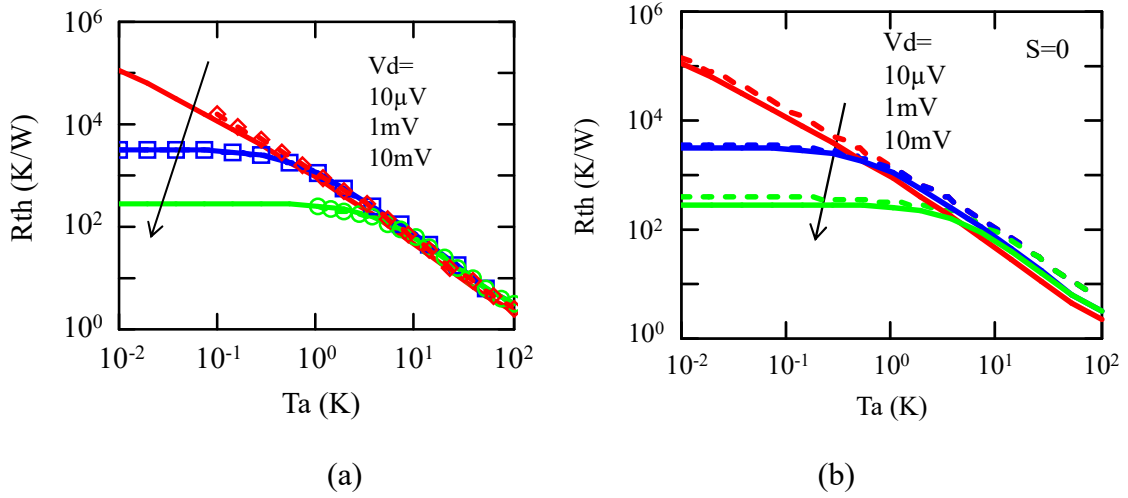

**Fig. 11.** a) Variations of thermal resistance $R_{th}$ with ambient temperature $T_a$ as obtained by numerical simulations without (solid lines) and with (symbols) consideration of thermopower in the electro-thermal transport equations for various drain voltages (long channel case FDSOI MOSFET, $t_{ox1}$=4nm, $t_{ox2}$=10nm, $t_{si}$=6nm, L=30nm, $V_{gl}$=1V, $T_S$=25K).

b) Variations of thermal resistance $R_{th}$ with ambient temperature $T_a$ as obtained by numerical simulations (solid lines) and with model $R_{thBSDG}(T_a,dT_{max})$ of Eq. (A6) (dashed lines) for various drain voltages (long channel case FDSOI MOSFET, $t_{ox1}$=4nm, $t_{ox2}$=10nm, $t_{si}$=6nm, L=30nm, $V_{gl}$=1V, $T_S$=25K).

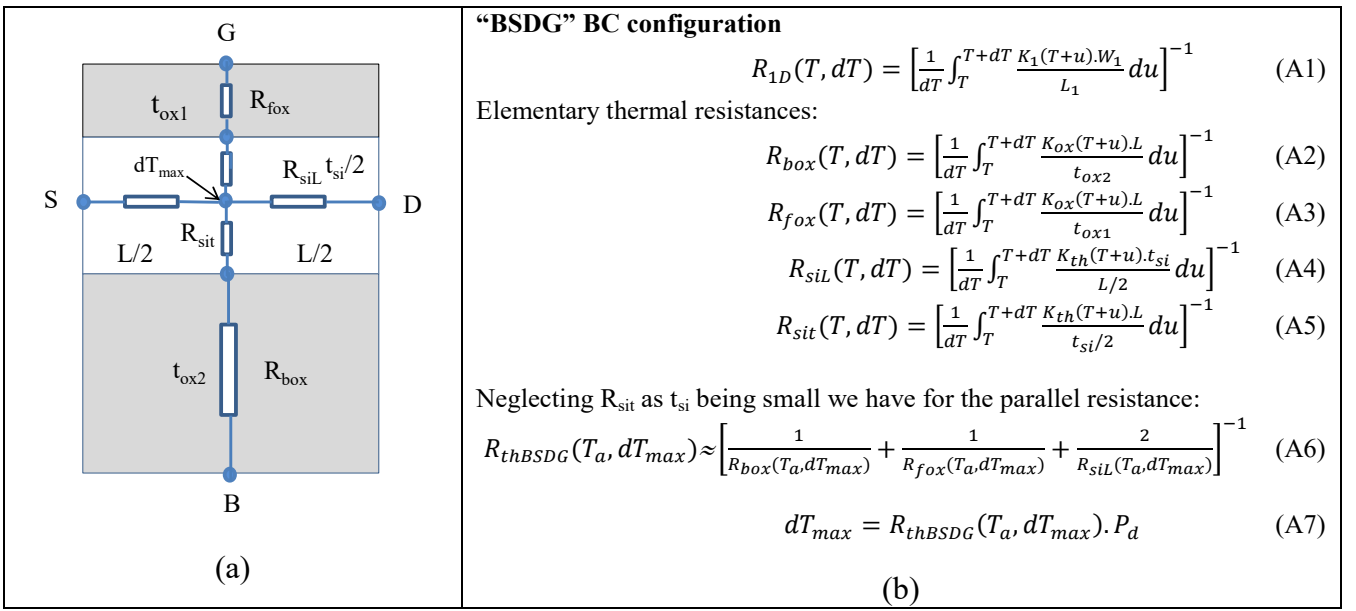

"BSDG" BC configuration

$$R_{1D}(T, dT) = \left[\frac{1}{dT}\int_T^{T+dT} \frac{K_1(T+u).W_1}{L_1} du\right]^{-1} \quad (A1)$$

Elementary thermal resistances:

$$R_{box}(T, dT) = \left[\frac{1}{dT}\int_T^{T+dT} \frac{K_{ox}(T+u).L}{t_{ox2}} du\right]^{-1} \quad (A2)$$

$$R_{fox}(T, dT) = \left[\frac{1}{dT}\int_T^{T+dT} \frac{K_{ox}(T+u).L}{t_{ox1}} du\right]^{-1} \quad (A3)$$

$$R_{siL}(T, dT) = \left[\frac{1}{dT}\int_T^{T+dT} \frac{K_{th}(T+u).t_{si}}{L/2} du\right]^{-1} \quad (A4)$$

$$R_{sit}(T, dT) = \left[\frac{1}{dT}\int_T^{T+dT} \frac{K_{th}(T+u).L}{t_{si}/2} du\right]^{-1} \quad (A5)$$

Neglecting $R_{sit}$ as $t_{si}$ being small we have for the parallel resistance:

$$R_{thBSDG}(T_a, dT_{max}) \approx \left[\frac{1}{R_{box}(T_a,dT_{max})} + \frac{1}{R_{fox}(T_a,dT_{max})} + \frac{2}{R_{siL}(T_a,dT_{max})}\right]^{-1} \quad (A6)$$

$$dT_{max} = R_{thBSDG}(T_a, dT_{max}).P_d \quad (A7)$$

(a)          (b)

**Fig. 12.** a) Schematic of thermal resistance equivalent circuit used for the analytical modelling of effective thermal resistance in the FDSOI structure for "BSDG" BC configuration. b) Formula of elementary thermal resistances in BOX, front oxide and silicon layer in the FDSOI structure.

The characteristics $dT_{max}(P_d)$ shown in Fig. 13 provide further insights into the behavior of self-heating. Both numerical simulations and analytical modeling using Eq. (A7) are compared, demonstrating excellent agreement between the two. From the figure, it can be observed that as the temperature is lowered, the self-heating temperature rise $dT_{max}$ more easily enters the sublinear regime with respect to power, leading to nonlinearity in the thermal resistance. Additionally, at high power in the sublinear regime and at low ambient temperature, the maximum temperature rise $dT_{max}$ typically varies as $\approx P_d^{0.5}$. This behavior can be attributed to the temperature dependence of $K_{ox}(T)$ and $K_{si}(T)$, which are involved in Eqs (A2) to (A6).

Figure 14 presents the variations of the thermal resistance $R_{th}$ with power as obtained from both numerical simulations and analytical modeling using Eqs (A6) and (A7). The very good agreement between the two results is noticeable, which confirms the physical validity of the analytical model. The asymptotic decrease of $R_{th}$ with power as $P_d^{-0.5}$ is directly derived from the $P_d^{0.5}$ dependence observed in the $dT_{max}(P_d)$ curves shown in Fig. 13. This behavior of $R_{th}$ with power can be attributed to the nonlinearity phenomenon resulting from the temperature dependence of $K_{ox}(T)$ and $K_{si}(T)$.

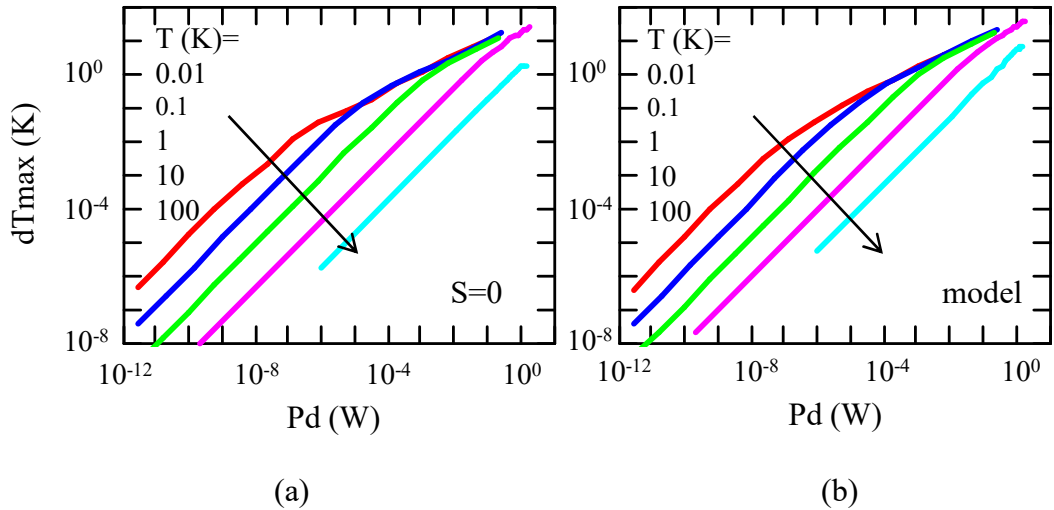

(a)          (b)

**Fig. 13.** Variations of maximum temperature rise $dT_{max}$ with power $P_d$ as obtained by (a) numerical simulations and (b) analytical model of Eq. (A7) for various ambient temperatures (long channel case FDSOI MOSFET, $t_{ox1}$=4nm, $t_{ox2}$=10nm, $t_{si}$=6nm, L=30nm, $V_{g1}$=1V, $V_d$=0 to 0.125V, $T_S$=25K).

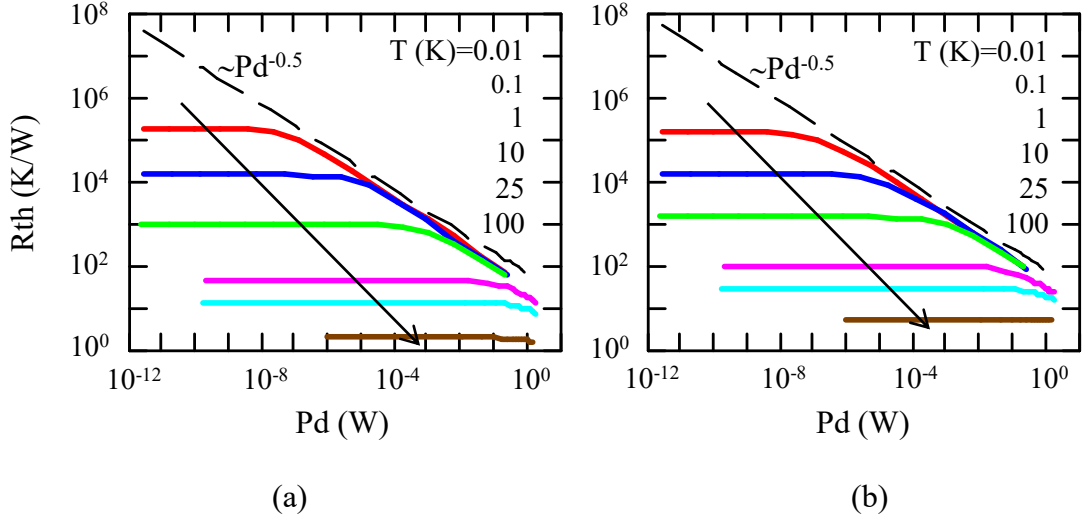

(a)          (b)

**Fig. 14.** Variations of thermal resistance $R_{th}$ with power $P_d$ as obtained by (a) numerical simulations and (b) analytical model of Eqs (A6) & (A7) for various ambient temperatures (long channel case FDSOI MOSFET, $t_{ox1}$=4nm, $t_{ox2}$=10nm, $t_{si}$=6nm, L=30nm, $V_{g1}$=1V, $V_d$=0 to 0.125V, $T_S$=25K).

The simulation results for the thermal resistance $R_{th}(T_a)$ in short-channel FDSOI MOSFETs with different boundary conditions ($V=V_{bi}$ at source and $V=V_{bi}+V_d$ at drain, see Fig. 3) were compared to those obtained in long-channel devices. The results, shown in Fig. 15, reveal that the $R_{th}(T_a)$ curves for short-channel devices closely resemble those for long-channel devices. This suggests that, despite the significant degradation of drain current in short devices due to the reduction in mobility caused by the saturation velocity effect, the thermal resistance remains largely unchanged at constant power. The reason for this observation is likely because the heat flows in the device are primarily governed by the phonon-originated thermal conductivities $K_{si}(T_a)$ and $K_{ox}(T_a)$, which are not significantly influenced by the electronic transport characteristics.

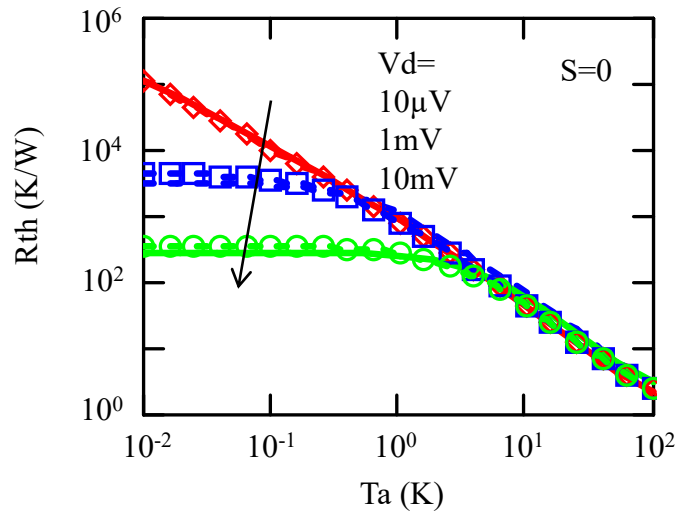

**Fig. 15.** Variations of thermal resistance $R_{th}$ with ambient temperature $T_a$ as obtained by numerical simulations for long (solid lines) and short (symbols) channel case FDSOI MOSFETs without consideration of thermopower in the electro-thermal transport equations for various drain voltages in short channel case. ($t_{ox1}$=4nm, $t_{ox2}$=10nm, $t_{si}$=6nm, L=30nm, $V_{gl}$=1V, $T_S$=25K).

### 3.2. Self-heating results for "B" BC configuration

Figure 16a shows the output drain current $I_d(V_d)$ characteristics obtained by numerical simulation with and without self-heating for various ambient temperatures in a long channel case FDSOI MOSFET with "B" BC configuration. The corresponding relative drain current changes due to self-heating, $dI_d/I_d$, are displayed in Fig. 16b. It is worth noting that the impact of self-heating on the drain current amplitude is much larger in this "B" BC configuration than in "BSDG" BC case (see Fig. 7). This is due to the fact that, as can be seen from Fig. 17 where are reported the $dT_{max}(P_d)$ and $R_{th}(P_d)$ characteristics for various temperatures, the temperature rise $dT_{max}$ and the thermal resistance $R_{th}$ are about 10 times larger than those found in the "BSDG" BC configuration (see Figs 9 and 10). It should be mentioned that when the simulated self-heating temperature rise $dT_{max}(P_d)$ is renormalized to the same device width, the amplitudes depicted in Fig. 17 reasonably align with experimental data obtained from real FDSOI MOS transistors [31]. This alignment strongly suggests that the primary heat flow in FDSOI devices is limited by the thermal resistance of the BOX (Buried Oxide) layer.

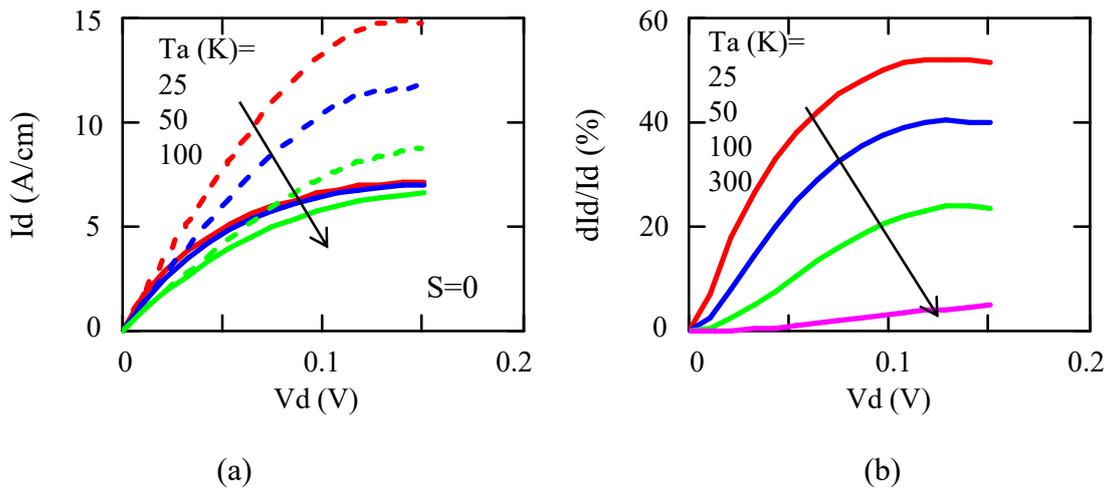

(a)            (b)

**Fig. 16.** a) Temperature dependence of (a) $I_d(V_d)$ output characteristics obtained by numerical simulations with (solid lines) and without (dashed lines) SHE.

(b) Corresponding relative drain current change $dI_d/I_d(V_d)$ due to SHE ("B" BC configuration, long channel case FDSOI MOSFET, $t_{ox1}$=4nm, $t_{ox2}$=10nm, $t_{si}$=6nm, L=30nm, $V_{gl}$=1V, $T_S$=25K).

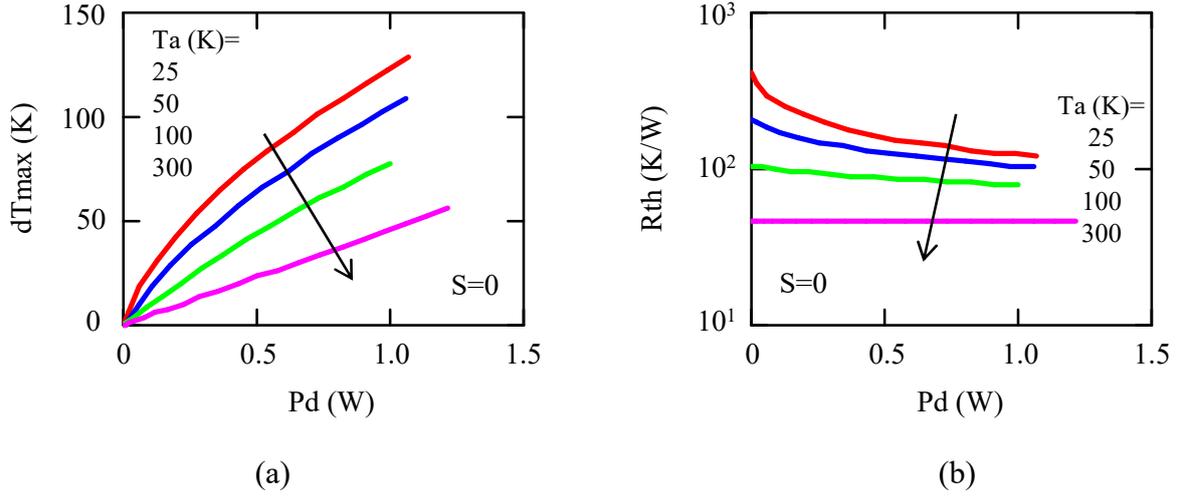

(a)　　　　　　　　　　　　　　(b)

**Fig. 17.** Temperature dependence of (a) $dT_{max}(P_d)$ characteristics and (b) $R_{th}(P_d)$ characteristics as obtained by numerical simulations without consideration of thermopower in the electro-thermal transport equations ("B" BC configuration, long channel case FDSOI MOSFET, $t_{ox1}$=4nm, $t_{ox2}$=10nm, $t_{si}$=6nm, L=30nm, $V_{g1}$=1V, $T_S$=25K).

In Fig. 18, the thermal resistance ($R_{th}$) variations with ambient temperature ($T_a$) are presented for different drain voltage ($V_d$) values in the "B" BC configuration of a long channel case FDSOI MOSFET. The thermal resistance values are obtained through numerical simulation and are represented by symbols on the graph. The behavior of the thermal resistance variations in the "B" BC configuration is similar to that observed in the "BSDG" BC configuration (see Fig. 11). However, the $R_{th}$ values in the "B" BC configuration are approximately ten times larger, as mentioned earlier. This disparity can be attributed to the fact that in this configuration, the heat flow is primarily limited by the thermal resistance of the BOX (Buried Oxide) layer. The graph provides an understanding of how the thermal resistance changes with ambient temperature for different drain voltages, highlighting the impact of the "B" BC configuration and the influence of the BOX thermal resistance on the heat dissipation in the FDSOI MOSFET.

In the "B" BC configuration, a similar approach is followed as in the "BSDG" BC case to develop an analytical model. The simplified thermal resistance equivalent circuit depicted in Fig. 19a is utilized in this configuration. In this case, the heat flow due to self-heating (SHE) is limited by the combined effect of $R_{sit}$ (silicon resistance) and $R_{box}$ (BOX thermal resistance). The temperature division between these resistances is expressed by Eq. (B3). The entire thermal resistance, denoted as $R_{thB}(T_a,dT_{max})$, is computed by solving Eqs (B3) and (B4) in a self-consistent manner. The variations of $R_{thB}(T_a,dT_{max})$ are then calculated using the $dT_{max}$ values obtained from the numerical simulations. These calculated variations are presented in Fig. 18 as solid lines. It is important to highlight that the analytical model and the simulation results exhibit good agreement, indicating that the model accurately captures the behavior of the thermal resistance. Additionally, the results suggest that $R_{thB}$ is primarily influenced by the BOX thermal resistance. Actually, $R_{thB}(T_a,dT_{max})$ can well be approximated by $R_{box}(T_a,dT_{max})$ since their difference is not exceeding 10% for ambient temperatures below 0.03K.

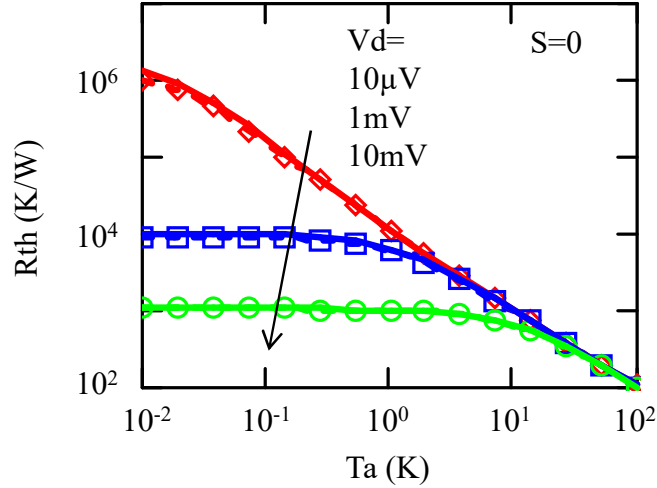

**Fig. 18.** Variations of thermal resistance $R_{th}$ with ambient temperature $T_a$ as obtained by numerical simulations (symbols) and with model $R_{thB}(T_a,dT_{max})$ of Eqs (B3) & (B4) (dashed lines) for various drain voltage values ("B" BC configuration, long channel case FDSOI MOSFET, $t_{ox1}$=4nm, $t_{ox2}$=10nm, $t_{si}$=6nm, L=30nm, $V_{g1}$=1V, $T_S$=25K).

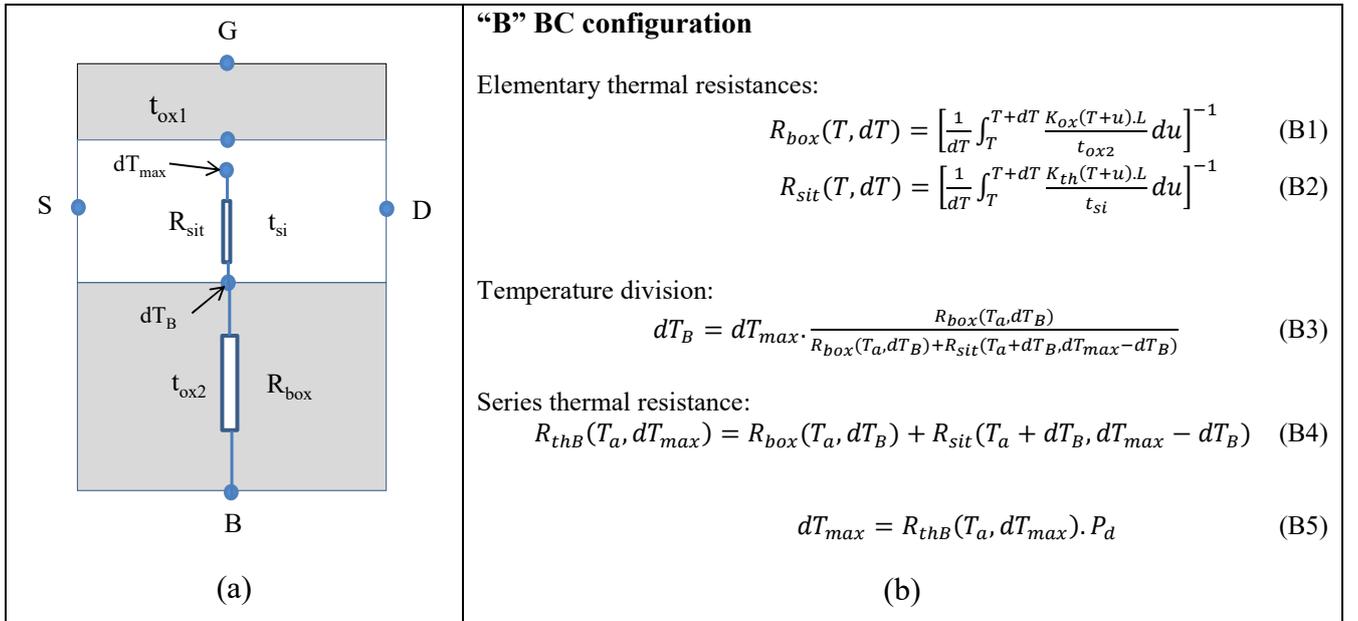

(a)            (b)

**Fig. 19.** a) Schematic of thermal resistance equivalent circuit used for the analytical modelling of effective thermal resistance in the FDSOI structure for "B" BC configuration. b) Formula of elementary thermal resistances in BOX and silicon layer in the FDSOI structure.

Figures 20 and 21 illustrate the self-heating performances obtained using the analytical models described in Eqs (A7) and (B5) for the FDSOI MOSFET device, considering very low ambient temperatures and dissipated powers. In Fig. 20, the variations of the self-heating temperature rise ($dT_{max}$) with the dissipated power ($P_d$) are compared between the "worst" and "best" thermalization cases, represented by the "B" and "BSDG" BC configurations, respectively. As previously mentioned, in the "worst" case where heat evacuation is primarily limited by the BOX thermal resistance, the self-heating temperature rise is approximately ten times larger than in the "best" case. Figure 21 displays the variations of the self-heating temperature rise ($dT_{max}$) and the device temperature ($T_{dev}=T_a+dT_{max}$) with ambient temperature for different dissipated powers in the "worst" thermalization case. The graph clearly indicates that as the ambient temperature decreases to very deep cryogenic levels, the device temperature ($T_{dev}$) becomes more influenced by the dissipated power. This highlights the critical importance of controlling the dissipated power for proper operation of FDSOI MOSFETs at extremely low cryogenic temperatures. These figures provide valuable insights into the self-heating behavior of the FDSOI MOSFETs under different thermalization scenarios, emphasizing the impact of the BC configuration and the necessity of managing dissipated power at deep cryogenic temperatures for reliable device performance.

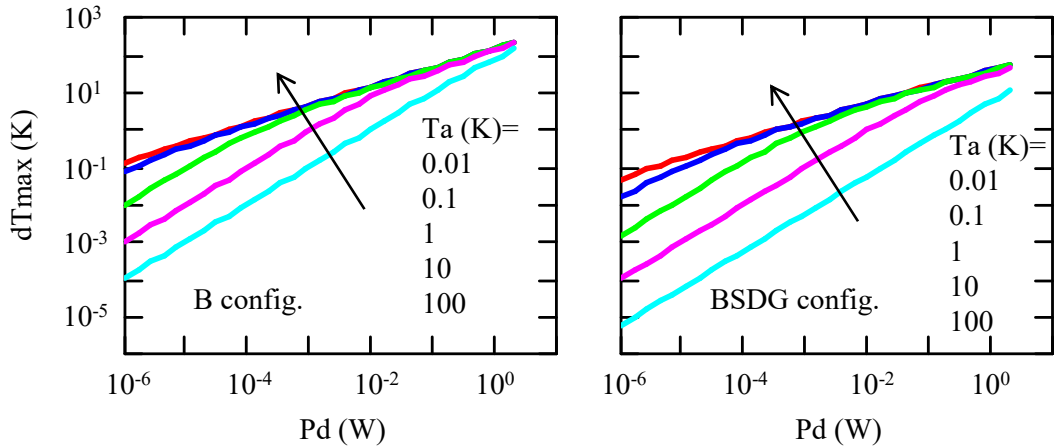

**Fig. 20.** Comparison of SHE temperature rise $dT_{max}$ variations with dissipated power $P_d$ as obtained from analytical modelling for "B" (left) and "BSDG" (right) BC configurations (FDSOI MOSFET, $t_{ox1}$=4nm, $t_{ox2}$=10nm, $t_{si}$=6nm, L=30nm).

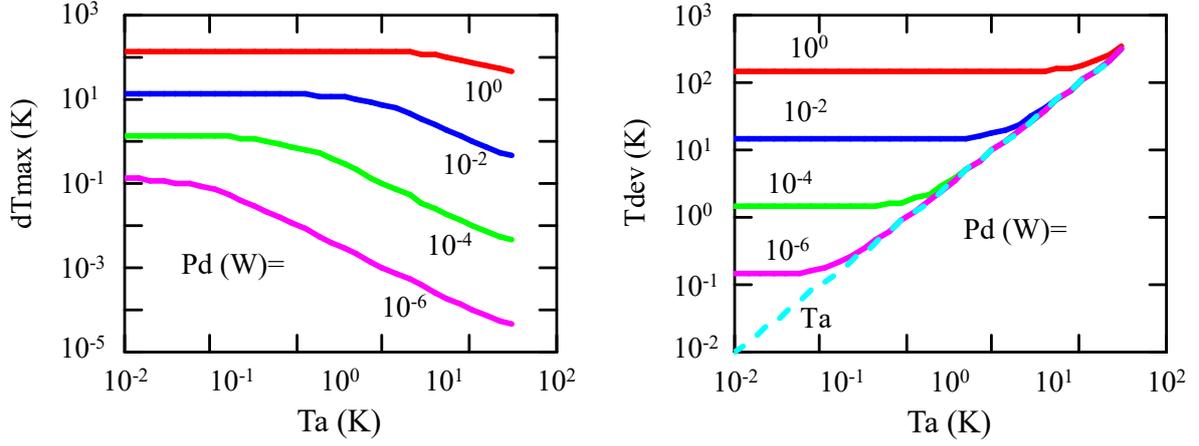

**Fig. 21.** Variations with ambient temperatures $T_a$ of SHE temperature rise $dT_{max}$ (left) and device temperature $T_{dev}=T_a+dT_{max}$ (right) for various dissipated power $P_d$ as obtained from analytical modelling for "B" BC configurations (FDSOI MOSFET, $t_{ox1}$=4nm, $t_{ox2}$=10nm, $t_{si}$=6nm, L=30nm).

## 4. Summary and conclusions

In this study, we have conducted TCAD numerical simulations to investigate the self-heating behavior of 30nm FDSOI MOS transistors at extremely low temperatures. This is the first time such simulations have been performed in this temperature regime. The aim was to evaluate the self-heating temperature rise ($dT_{max}$) and thermal resistance ($R_{th}$) as functions of ambient temperature ($T_a$) and dissipated electrical power ($P_d = I_d.V_d$), using calibrated values for the thermal conductivities of silicon and oxide. The numerical simulations revealed several important findings. Firstly, the characteristics of the SHE temperature rise, $dT_{max}(P_d)$, displayed a sub-linear behavior at sufficiently high dissipated powers, which is consistent with previous experimental results reported in the literature. Additionally, the simulations demonstrated that at very deep cryogenic temperatures, the SHE temperature rise ($dT_{max}$) can significantly surpass the ambient temperature more easily. Furthermore, a detailed thermal analysis of the heat flows within the FDSOI structure, considering the non-linear thermal resistances in the silicon channel and front/BOX oxides, was conducted. Based on this analysis and the TCAD simulation data, an analytical model for self-heating effects was developed and calibrated. This analytical model successfully describes the temperature rise due to self-heating and the thermal resistance as function of dissipated power in FDSOI MOS devices operating at ambient temperatures as low as 10mK. Moreover, it is worth noting that this self-heating model physically justifies the simple analytical modelling approach developed in [32]. Consequently, these TCAD simulations and the analytical modelling provide valuable insights into the self-heating and electro-thermal performance of FDSOI MOS transistors, considering the interplay of ambient temperature and dissipated power. These findings can be particularly useful for cryogenic circuit design, enabling the prediction and optimization of self-heating effects in FDSOI MOS devices.

## 5. Acknowledgment

The authors are grateful to EU H2020 RIA project SEQUENCE (Grant No. 871764) for financial support.

***************